\documentclass[%
 aip,
 amsmath,amssymb,
 reprint,%
]{revtex4-1}
\usepackage[none]{hyphenat}
\usepackage{amsmath}
\usepackage{amssymb}
\usepackage{amsfonts}
\usepackage{array}
\usepackage{graphicx}
\usepackage{mathrsfs}
\usepackage{multirow}
\usepackage{siunitx}
\usepackage{subcaption}
\usepackage{capt-of}
\usepackage{soul}
\usepackage{xcolor}

\newcolumntype{M}[1]{>{\centering\arraybackslash}m{#1}}

\draft 

\makeatletter
\def\@email#1#2{%
 \endgroup
 \patchcmd{\titleblock@produce}
  {\frontmatter@RRAPformat}
  {\frontmatter@RRAPformat{\produce@RRAP{*#1\href{mailto:#2}{#2}}}\frontmatter@RRAPformat}
  {}{}
}%
\makeatother

\begin{document}

\newtheorem{theorem}{Theorem}

\title{Experimental and simulative study on laser irradiation of 3D-printed micro-structures at intensities relevant for inertial confinement fusion}

\author{M. Cipriani}
\email{mattia.cipriani@enea.it}

\author{F. Consoli}
\author{M. Scisci\'o}
\affiliation{ENEA, Nuclear Department, C.R. Frascati, Italy}

\author{A. Solovjovas}
\author{I. A. Petsi}
\author{M. Malinauskas}
\affiliation{Laser Nanophotonics Group, Laser Research Center, Physics Faculty, Vilnius University, Vilnius, Lithuania}

\author{P. Andreoli}
\author{G. Cristofari}
\author{E. Di Ferdinando}
\author{G. Di Giorgio}
\affiliation{ENEA, Nuclear Department, C.R. Frascati, Italy}


\begin{abstract}
Inertial confinement fusion requires a constant search for the most effective materials for improving the efficiency of the compression of the capsule and of the laser-to-target energy transfer. 
Foams could provide a solution to these problems, but they require further experimental and theoretical investigation. 
The new 3D-printing technologies, such as the two-photon polymerization, are opening a new era in the production of foams, allowing for the fine control of the material morphology. 
Detailed studies of their interaction with high-power lasers in regimes relevant for inertial confinement fusion are very few in the literature so far and more investigation is needed. 
In this work we present the results an experimental campaign performed at the ABC laser facility in ENEA Centro Ricerche Frascati where 3D-printed micro-structured materials were irradiated at high power. 
3D simulations of the laser-target interaction performed with the FLASH code reveal a strong scattering when the center of the focal spot is on the through hole of the structure.
The time required for the laser to completely ablate the structure obtained by the simulations is in good agreement with the experimental measurement.
The measure of the reflected and transmitted laser light indicates that the scattering occurred during the irradiation, in accordance with the simulations.
Two-plasmon decay has also been found to be active during irradiation.
\end{abstract}


\maketitle

\section{Introduction}\label{sec:introduction}

The peculiar way high-power lasers interact with micro-structured materials, or foams, and the evolution of the plasma generated by the laser action proved to be useful for various kind of applications. 
They were used as efficient sources of X-rays \cite{Perez2014}, for the study of shock waves \cite{Benuzzi1998}, equations of state \cite{Batani2001} and for efficient particle acceleration, neutron and betatron generation with short laser pulses \cite{Rosmej2020}.
In particular, they find many applications in Inertial Confinement Fusion (ICF), because of their ability to enhance laser absorption \cite{Cipriani2021}, to smooth laser inhomogeneities in the transverse energy profile \cite{Depierreux2009} and to enhance the ablation loading on a substrate \cite{DeAngelis2015}.
Recently, low-density foams were seen to reduce the SBS instability, which has been linked to an increased ratio between the ion and the electron temperature \cite{Mariscal2021, Hudec2025}.
They have been suggested to be employed inside the hohlraums to prevent wall expansion \cite{Iaquinta2024,Moore2020} and to function as support material for liquid nuclear fuel, named as wetted foams \cite{Olson2021,Igumenshchev2023}.
New recent advances in manufacturing, simulations and diagnostics are opening the way to a deeper and more fundamental investigation of the characteristics of the plasma generated by the action of high-power lasers.

Until a few years ago, the way of manufacturing foams of light elements was through chemical processes, creating a gel and then removing the solvent from it (see for example \cite{Nagai2018} for a review on the subject).
We refer to the foams produced in this way as \textit{chemical} foams.
These foams are convenient because of the low price per sample and because they can be realized in a large number in a short time.
On the other hand, the structural parameters, such as the pore size, the solid part thickness and the average density, can be controlled with precision only in some ranges of values, and they may vary from sample to sample because of the intrinsic stochasticity of the production process.
Moreover, the manipulation of these materials during manufacturing and handling in experiments poses often serious difficulties, due to the softness and fragility of the structure, especially at densities lower than 10 mg/cm$^3$.
Nowadays, the Laser Direct Writing (LDW) Multi-Photon 3D Lithography (MP3DL) technology based on two-photon polymerization is (see, for example, \onlinecite{23AFM2214211Two-Photon,24NMRP} for extensive review and tutorials) allows to obtain innovative foam samples with a predictable and reproducible structure, useful for laser-matter experiments \cite{Wiste2023, Jones2021}.
The technique is exceptionally attractive for rapid prototyping high-resolution complex-architectures for experimental research requiring advanced material engineering~\cite{24CM3obtaining}. 
Furthermore, recent developments towards increasing the throughput up to 10$^7$ voxels/s give very promising results for scaling up the production rate acceptable to mass-customized manufacturing~\cite{24OEA240003acousto}.
In particular, LDW allows for the precise design of the morphology of the sample, which can be in principle engineered for any experimental need, improving its performances but also the diagnostics of the laser-produced plasma.
Producing a rather limited amount of high-quality samples still requires several hours, which reflects in a high pricing per sample.
Also, the lower limit of the density is still close to 10 mg/cm$^3$ with empty spaces of the order of 10 µm \cite{Wiste2023}, which at the moment limits the possible applications.
Nonetheless, this technology is available from a few years and it has a wide margin for new developments which could bring these materials to the parameters needed for the aforementioned applications.

The simulation of the interaction of a high-power laser beam with such materials is a challenging task, due to the different scales of the foam morphology.
For chemical foams, the filament thickness ranges from 10s to 100s of nanometers, while the pore size goes from a few to tens of micrometers.
The stochasticity of the structure adds even more difficulties to the modeling.
To computationally resolve this problem, several sub-grid models have been implemented in radiation-hydrodynamic codes \cite{Velechovsky2016, Cipriani2018c, Belyaev2020, Hudec2023}.
The advantage of using these models resides in the fact that their implementation in the codes does not appreciably change the computational performance of the original code.
On the other hand, the models necessarily approximate the physics involved in the plasma evolution and can lack accuracy. 
Simulations exploiting a more fundamental description of the laser-matter interaction would be preferable.
At present, the technology allows to follow this route. 
On the one hand, printed foams can have regular structures which can be in principle replicated in numerical codes with a one-to-one correspondence.
On the other hand, the development of parallel radiation-hydrodynamics codes and the advances in High Performance Computing (HPC) infrastructures enables a growing computational power, allowing to run more demanding simulations in a reduced time \cite{Milovich2021}.

In this work we present the results of the first experimental campaign performed at the ABC laser facility in the ENEA Research Center in Frascati for high-power irradiation of additively manufactured materials.
We used tailored micro-structured samples with a log-pile structure obtained through LDW with the SZ2080\texttrademark\  hybrid pre-polymer, supported by a holder 3D-printed with UV stereolithography, specifically designed to allow optimal access to the diagnostics fielded in the experiments.
All structural parameters, apart from the thickness, were the same across the whole set of samples, to ensure the highest possible shot-to-shot reproducibility.

We irradiated these samples at intensities relevant for ICF research, from $10^{14}$ W/cm$^2$ to about $10^{15}$ W/cm$^2$, to investigate the behavior of the laser-generated plasma and in particular the speed at which the material was eroded by the laser action.
We performed 3D radiation-hydrodynamics simulations with the FLASH code, in which the target micro-structure was directly reproduced, exploiting the parallelization capabilities of the code on the HPC cluster ENEA-CRESCO6 \cite{9188135}.
The filament thickness and their separation in the targets used in the experiment were chosen to lower the computational cost for 3D simulations, having a more coarse structure.
This lengthened the time required to fill the gaps between the filaments, thus enhancing the effects of some physical processes, such as the laser scattering in the sample, which would otherwise be less evident.


\section{The targets}\label{sec:targets}
\begin{figure}[tbp]
    \centering
    \begin{subfigure}[b]{0.48\textwidth}
        \includegraphics[width=\linewidth]{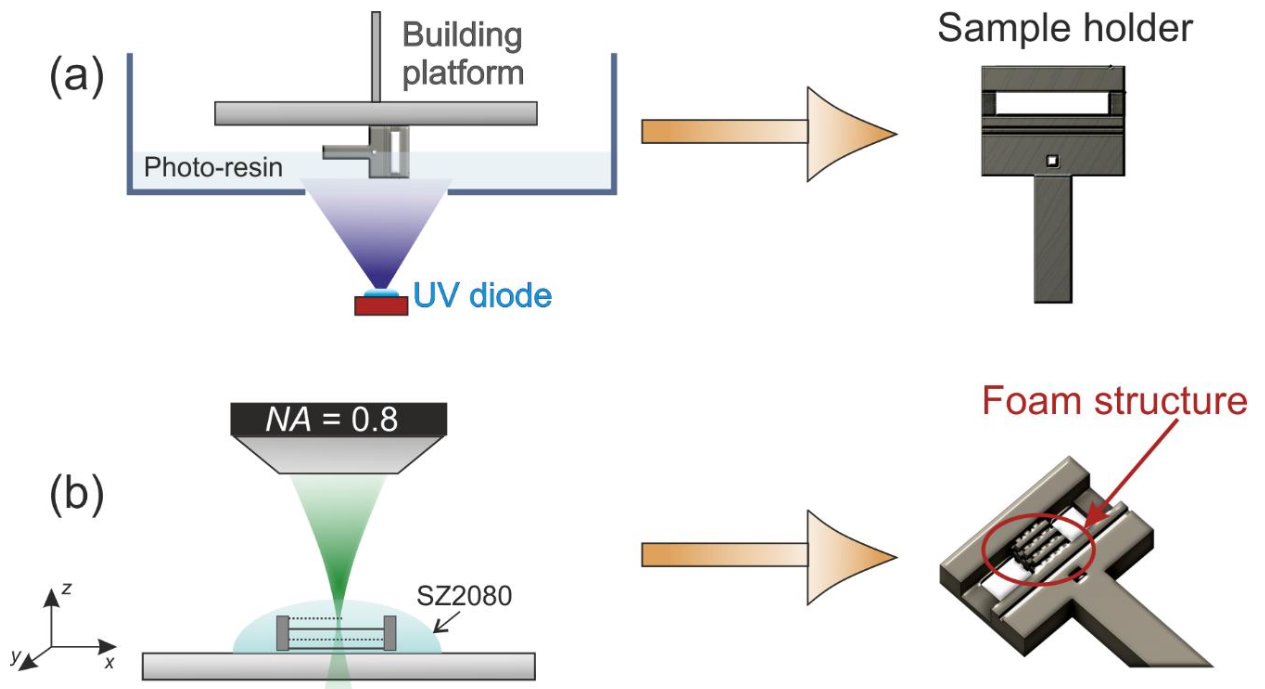}
        \caption*{}\label{fig:sample-production}
    \end{subfigure}
    \begin{subfigure}[b]{0.23\textwidth}
        \addtocounter{subfigure}{2}
        \includegraphics[width=\textwidth]{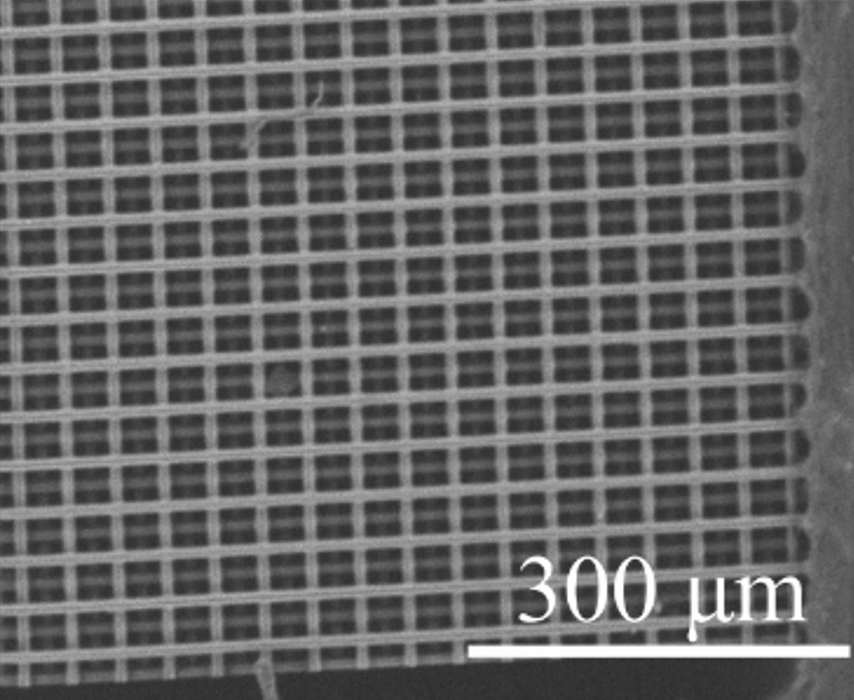}
        \caption{}\label{fig:logpiles250}
    \end{subfigure}
    \begin{subfigure}[b]{0.23\textwidth}
        \includegraphics[width=\textwidth]{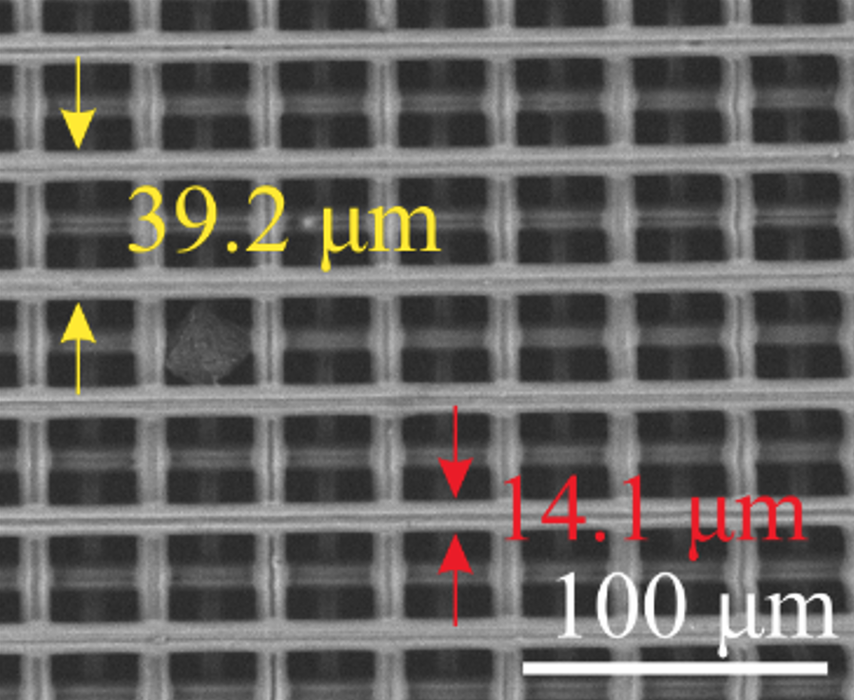}
        \caption{}\label{fig:logpiles600}
    \end{subfigure}
    
    \caption{An illustration of the techniques of production of the target holders (a) and of the micro-structures (b). Scanning electron microscopy images of the samples used in the experiments with 250x (\subref{fig:logpiles250}) and 600x (\subref{fig:logpiles600}) magnifications. In (\subref{fig:logpiles600}) the filament separation is indicated in yellow and the filament thickness in red. }
    \label{fig:sample-production}
\end{figure}

The samples used in the experiments were constituted by two parts, printed with different techniques, as shown in Figure \ref{fig:sample-production}.
The holder was realized with table-top 3D UV printer and Asiga Plasgray photo-resin to sustain the log-piles during manufacturing, avoiding deformations of the structures, and for handling in the experimental chamber during the irradiations.
The shape of the holder was designed to allow an optimal access to the diagnostics used in the experiment, described in Section \ref{sec:the-experiment}.
The micro-structures were made with a LDW MPL3DL printer directly into the holders, using the hybrid pre-polymer SZ2080\texttrademark\ .
This pre-polymer contains Si and Zr, with the chemical formula C$_4$H$_{12}$SiZrO$_2$.
We estimated that in the final structure the percentage in mass of the Si atoms was about $ 15\%$, while the one for the Zr atoms was about $20\%$.
The separation between the filaments was 39.2 µm and the filament thickness was 14.1~µm over the whole structure, with an average density of 348 mg/cm$^3$.
To increase the cross-section with the main laser beam under irradiation, we printed the layers in the log-pile with a shift of 1/2 of the period of the structure in both directions. 
The each sample had a square transverse shape with sides of 500 µm and a variable thickness from about 100 µm to 400 µm.
Additional details of the fabrication process can be found in the Appendix \ref{app:manufacturing}.


\section{The experiment}\label{sec:the-experiment}

\begin{figure}[tbp]
    \centering
    \includegraphics[width=\linewidth]{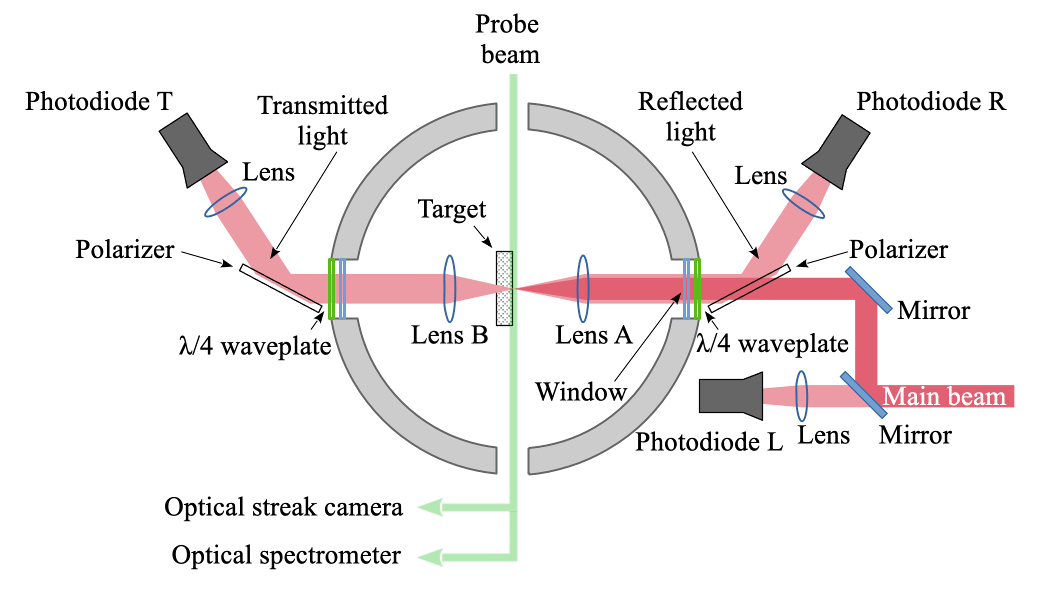}
     \caption{The experimental setup used in the campaign. 
     The pulse temporal profile was monitored by the fast Photodiode L, collecting a leak from a mirror on the laser optical path.
     The laser light reflected and transmitted by the target was collected by the focalization lenses and detected by the two fast Photodiodes R and L. 
     A visible streak camera imaged the light self-emitted by the plasma at 90° from the laser propagation direction.
     At the same position a visible spectrometer was used to spectrally resolve the plasma self-emission.}
    \label{fig:setup}
\end{figure}

The results discussed in this work regard a recent experimental campaign carried out at the ABC laser facility at the ENEA Centro Ricerche Frascati.
ABC is a Nd:glass laser able to deliver two counter-propagating beams with a wavelength of 1054 nm and a maximum energy of 100 J each. 
For this experiment we used a single beam configuration in a planar irradiation geometry.
The temporal shape of the pulse was a $\sin^2 t$ with a full duration of 5 ns, 3 ns full width half maximum, with a total energy of about 40 J.
The pulse energy was maintained as constant as possible among the different shots, in order to assess the repeatability of the experimental results.
The beam was focused on the target at normal incidence by using an $f$/1 lens.
Two focal spot diameters were used, namely 50 µm and 100 µm, resulting in a lower and a maximum intensity of $I_{\text{lower}}1.3 \times 10^{14}$ W/cm$^2$ and $I_\text{max} = 7 \times 10^{14}$ W/cm$^2$, respectively.
The laser energy in the spot had a gaussian distribution.

A sketch of the experimental setup is shown in Figure \ref{fig:setup}.
The laser temporal profile was monitored for each shot by a calibrated photodiode (Photodiode L).
The laser light reflected by and transmitted through the target were collected by the two focusing lenses A and B with the following method. 
The main beam light was linearly polarized, passed through a polarizer which as the same orientation and entered the interaction chamber through a $\lambda/4$ plate, which changed the polarization from linear to circular. 
The light reflected by the target and intercepting the focusing lens exited the chamber passing through the same $\lambda/4$ plate and its circular polarization was turned into linear again, but in an orthogonal direction with respect to the incoming laser light, being thus rejected by the polarizer and collected by the Photodiode R.
On the other side of the chamber, a mirror version of this setup was in place and the Photodiode T then collected the laser light transmitted though the target.
A green probe beam with a wavelength of 527 nm and a time duration of 500 ps was generated by picking a small portion of the main beam light along the amplification path and frequency-doubling it, so that it was absolutely synchronized with the main beam.
The probe beam was delayed with respect to the main beam by 12 ns.
A Hamamatsu C5680 visible streak camera watched the target from the side, to record the self-emitted light from the plasma generated at the front and at the back of the target.
The probe beam reached the slit of the streak camera, thus acting as a time fiducial on the streaked image.
A Ocean Optics HR4000 spectrometer collected the self-emitted light from the plasma in the visible range in a time-integrated manner.

\section{Simulation results}\label{sec:simulations}

Before discussing the experimental outcomes, we show the results of the numerical simulations, which clarify the hydrodynamic behavior of the material under irradiation, in particular regarding the degree of homogenization of the plasma.
To simulate the behavior of the material under irradiation we used the FLASH code \cite{Fryxell2000, Dubey2009}.
FLASH is an open-source, modular, multi-physics code in which the plasma fluid equations are solved with finite-volume methods on an Eulerian grid. The code has Adaptive Mesh Refinement (AMR) capabilities, implicit solvers for diffusion, laser ray-tracing, radiation diffusion, and multi-material support with tabulated Equations of State (EoS) and opacities.
All the simulations were run on the ENEA CRESCO high performance computing infrastructure \cite{9188135}.
The large filament thickness and inter-filament spacing allowed to directly reproduce the sample structure in the code as the initial condition and complete the simulation in a reasonable time, therefore avoiding approximations or specific modeling for describing the plasma evolution.
The mesh has been chosen to be static rectangular with a cell size of about 1 µm in each spatial direction.
Such resolution allowed us to properly resolve the shock dynamics inside the filaments.
We did not exploit the AMR capabilities of the FLASH code, since the refinement process of such a large number of cells would have impacted performance.
In fact, the very inhomogeneous plasma quickly filling almost the whole domain from the very beginning of the simulation would have required the most refined mesh everywhere, thus not improving performance compared to a static mesh.
The 3D ray tracing available in the FLASH code allowed to reproduce the angular incidence of the ABC laser beam on the sample and its scattering on the filaments at the beginning of the interaction, before the closure of the empty spaces due to the plasma fill.
The density of the filaments was set to be 1.2 g/cm$^3$, in accordance with the density of the printed material.
As discussed in Section \ref{sec:targets}, the printed structures contained Si and Zr and the role played by radiation must be in principle accounted for in the simulations.
Considering the estimated percentages in mass of the two elements, we computed the opacities from the TOPS website \cite{zotero-89854}.
Due to the high computational cost of including the radiation transport, we compared a simulation with radiation transport another without it, to assess to which extent this played a role in the plasma evolution in our specific experimental situation.
The differences between the two cases were very small in terms of the speed of the erosion wave, which was our main experimental observable, as discussed in Section \ref{sec:exp-results}.
Therefore, we ran all the simulations reported in this work without radiation, strongly reducing the computational cost and allowing us to get the results in a reasonable time. 

\begin{figure}[tbp]
        \includegraphics[width=0.48\textwidth]{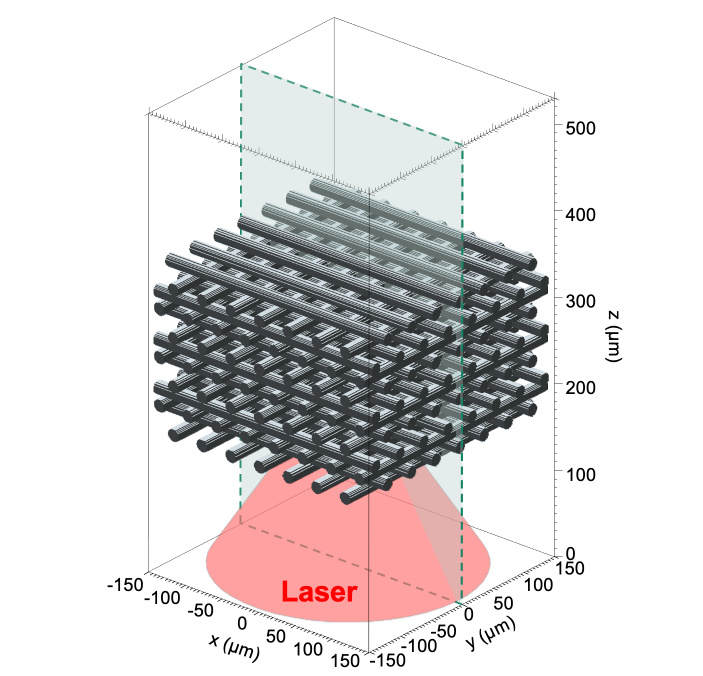}
    \caption{The initial configuration of the simulation. 
    The domain was 300 µm wide along the $x$ and $y$ axis and 530 µm wide along the $z$ axis.
    The plane along which the domain has been cut to show the results of the simulations presented is shown with a teal fill and a dashed contour.}
    \label{fig:sim-initial}
\end{figure}

\begin{figure}[tbp]
\centering
\begin{subfigure}[b]{0.15\textwidth}
    \includegraphics[width=\textwidth]{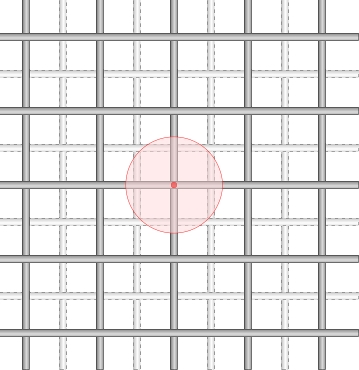}
    \caption{Case 50C}
    \label{fig:focusing-cases-50-cross}
\end{subfigure} \hspace{10pt} 
\begin{subfigure}[b]{0.15\textwidth}
    \includegraphics[width=\textwidth]{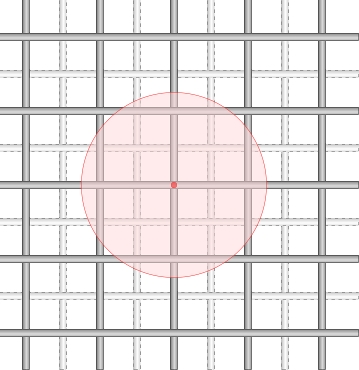}
    \caption{Case 100C}
    \label{fig:focusing-cases-100-cross}
\end{subfigure} \\
\begin{subfigure}[b]{0.15\textwidth}
    \includegraphics[width=\textwidth]{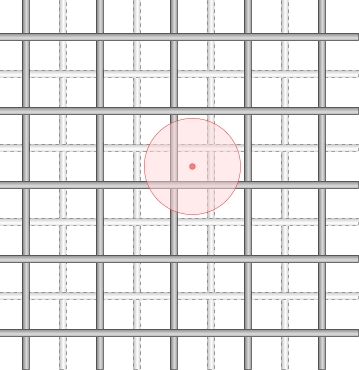}
    \caption{Case 50H}
    \label{fig:focusing-cases-50-hole}
\end{subfigure} \hspace{10pt}
\begin{subfigure}[b]{0.15\textwidth}
    \includegraphics[width=\textwidth]{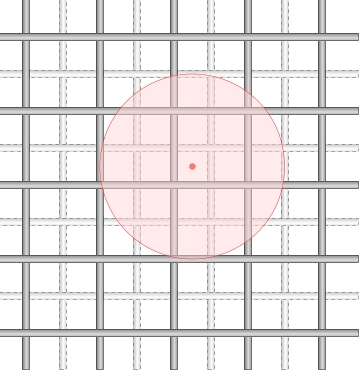}
    \caption{Case 100H}
    \label{fig:focusing-cases-100-hole}
\end{subfigure}
    \caption{The different cases of focusing.
    The light red circle indicates the focal spot, while the small red circle indicates its center.
    The labels correspond to the following cases: 50C, 50 µm diameter spot with the center on the crossing of filaments; 100C, the same as 50C but with a spot diameter of 100 µm; 50H, 50 µm diameter spot with the center on the hole between the filaments of all layers; 100H, the same as 50H but with a spot diameter of 100 µm.}
    \label{fig:focusing-cases}
\end{figure}

The initial setup of the simulation is shown in Figure \ref{fig:sim-initial}. 
We used the two focal spot diameters mentioned in Section \ref{sec:the-experiment}, namely 50~µm and 100~µm.
The 50~µm diameter was of the same order of the separation between the filaments in each single layer, while the 100~µm diameter could completely cover a square of four elementary cells.
For each focal spot size, we can identify two limit cases in terms of the position of the center of the focal spot, as depicted in Figure \ref{fig:focusing-cases}.
Since we used an $f$/1 lens in the experiments, distinguishing between these two cases has an important relevance, as we will see shortly.
We will refer to each of the cases, from here on, with the following labels: 100C, 100 µm diameter spot with the center on the crossing of filaments of the first layer; 50C, the same as 100C but with a spot diameter of 50 µm; 100H, 100 µm diameter spot with the center on the through hole between the filaments of all layers; 50H, the same as 100H but with a spot diameter of 50 µm.
A third pointing possibility can be identified, with the center of the focal spot placed on the crossing of the filaments on the second layer.
However, this pointing case is analogous to the cases C already discussed.

\begin{figure*}[tbp]
    \includegraphics[width=0.9\textwidth]{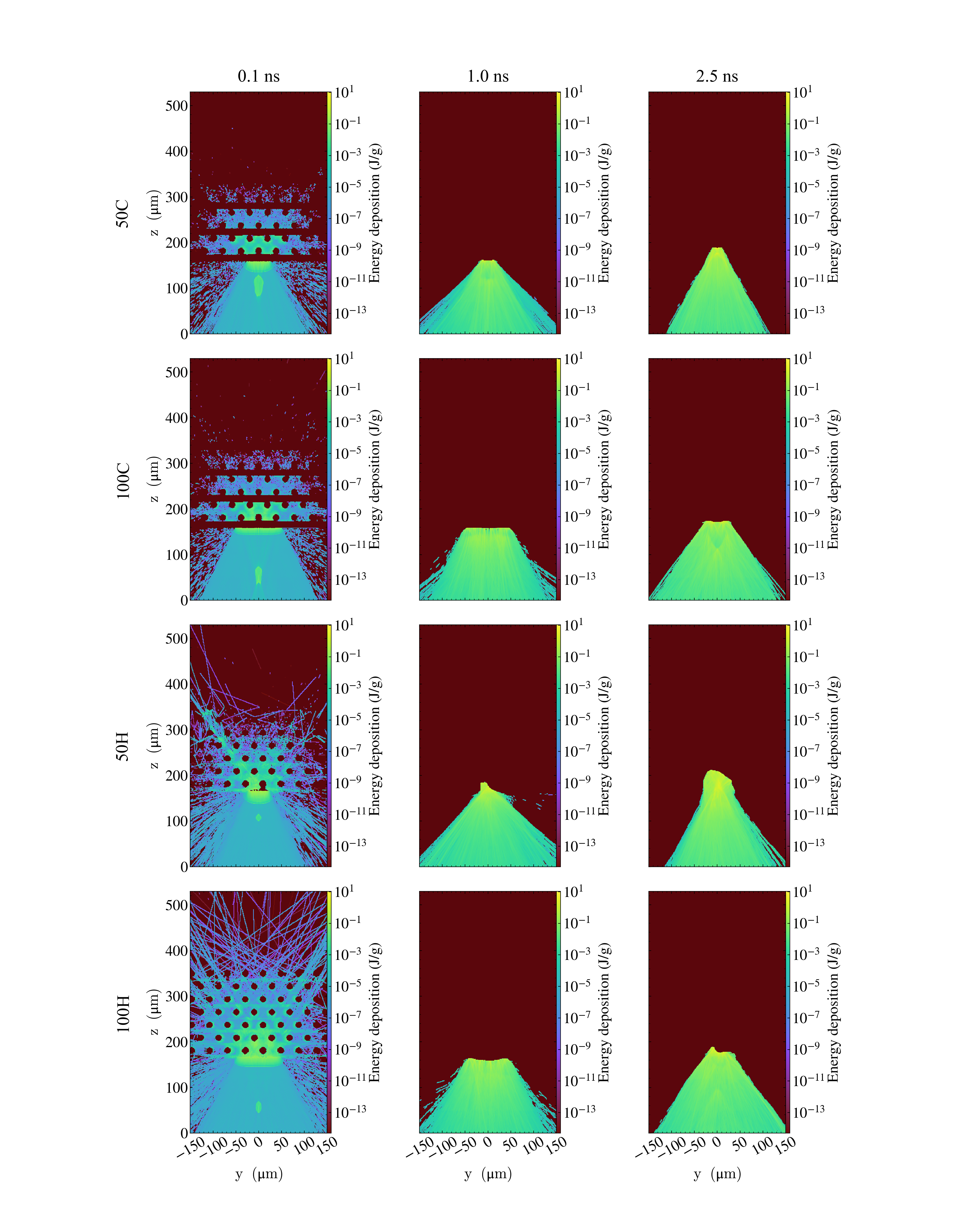}
    \caption{\label{fig:scattering} The time evolution of laser propagation inside the micro-structure. 
        The columns show from left to right the times 0.1 ns, 1 ns, 2.5 ns from the beginning of laser irradiation.
        For each time, the effect of different focusing of the laser is shown, from top to bottom, for the cases 50C, 100C, 50H, 100H.}
\end{figure*}

\begin{figure*}[tbp]
    \includegraphics[width=0.95\textwidth]{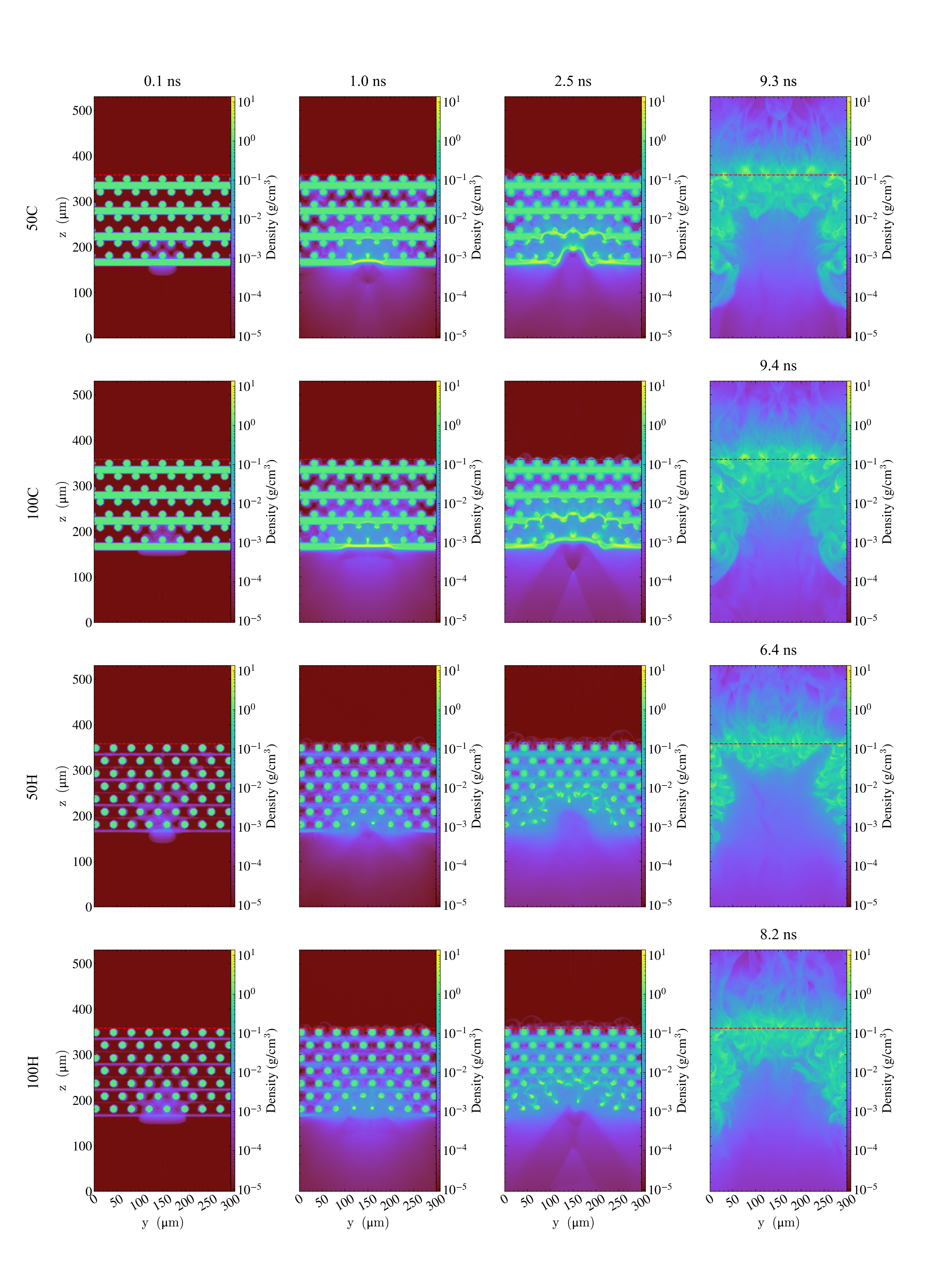}
    \caption{\label{fig:density-results} 
    The density plots at different times, depending on the focusing point on the structure.
    The red dashed line indicates the initial position of the back side of the target.}
\end{figure*}

We performed one simulation for each case, to asses the importance of the pointing of the laser on the micro-structure.
The simulations 50H and 100H were initialized as the 50C and 100C simulations, but with shifting the target by a quarter of the period of the structure in both $x$ and $y$ directions, keeping the laser direction along the central axis of the numerical domain.
Figure \ref{fig:scattering} reports the plots of the laser energy deposition, revealing the path of the laser rays as they traverse the micro-structure after 100 ps, 1 ns and 2.5 ns, the latter corresponding to the time at which the laser pulse reached its maximum power.
All the pictures in this Figure and in Figure \ref{fig:density-results} were obtained by slicing the numerical domain with a plane passing through the axis of the laser and normal to the $x$ axis.

In the cases of 50H and 100H a large amount of laser scattering into the structure is seen. 
The large focusing angle of the $f/1$ lens makes a part of the beam to impinge on the target at an angle which facilitates the scattering in a specific direction in the micro-structure, forming a channel, as clearly seen for the 50H case.
The heating resulting from this scattering is volumetric and affects all the layers of the material.
The scattering occurs more broadly realized in the 100H case because of the larger number of filaments and holes covered by the focal spot.
At later times, as can be seen in the pictures for the 50H case at 1 ns and for the 100H case at 3 ns in Figure \ref{fig:scattering}, zones of highly inhomogeneous plasma density appear, leading to the refraction of the laser rays into a small volume, which may result in self-focusing.
In fact, as discussed below, the expansion and ablation of the thick filaments leaves zones of high density in the plasma until the end of the simulation at 10 ns.
This may be the cause of the strong two-plasmon decay signature we observed in the experiments, see Section \ref{sec:LPIs} below.
In the cases of focusing on the crossing between filaments, namely 100C and 50C, the scattering is less evident and the heating depth in the target is similar.
In these cases, the geometrical factors which favor the scattering in the 50H and 100H cases are not sufficently well met and the laser is not scattered into channels.
The time required to the plasma to fill the gaps between the filaments, thus preventing the laser to penetrate the structure, was of about 500 ps in all cases.

Figure \ref{fig:density-results} shows the time evolution of the density of the target material and of the laser-generated plasma.
We see that after 1 ns in the 50C and 100C cases the target structure is still almost intact, with shock waves propagating into the filaments of the first two layers.
In these cases cases we can also observe that the centers of the filaments of the first two layers are slightly displaced from their original position in correspondence of the laser focal spot.
This displacement is evident at 2.5 ns in the 50H case at the peak of the laser power, where the two front layers have been completely ablated and displaced by the laser action.
In the 100H case, due to the lower intensity, this process is slower and the first two layers are starting to move and to pack towards the rear side of the target.
A different picture is seen in the cases 100C and 50C, where at 2.5 ns we observe a lower displacement of the filaments, due to the relatively shorter penetration of the laser into the structure, as observed in Figure \ref{fig:scattering}.
The motion of the heated filaments appears to be more symmetric in the C cases compared to the H cases.
At the end of the laser pulse at 5 ns the evolution of the material for the 100C and 50C cases is similar, also considering the depth at which the target has been eroded by the laser action.
From the images in Figure \ref{fig:density-results} we can readily see that during the laser pulse only a partial homogenization of the plasma is achieved. 
As discussed in Section \ref{sec:introduction}, this was expected considering the very large thickness of the filaments and their large separation.
Therefore, we cannot speak about a shock wave being generated in the material and we will refer to the propagation of the front of the region where the micro-structure has been ablated as an \textit{erosion wave}.
From the simulation results, we estimated the breakout time of the erosion wave at the rear side of the target by the following procedure.
We extracted the profile of the density along a line parallel to the $y$ axis and passing through the point $(x,y,z)=(0, 0, 370) \  \text{µm}$.
The value of the $z$ coordinate corresponds to the position of the rear side of the target at $t=0$ plus the diameter of a filament.
We then computed the average of the density along this line and we identified the time at which this value reached a maximum.
This corresponds to determining when the remnants of the filaments at the back of the target moved out of the target by a one layer distance.
The time identified by this criterion was defined as the erosion wave breakout time.
As it is seen in the last column of Figure \ref{fig:density-results}, the breakout time of the erosion wave happens at about the same time for both 50 µm and 100 µm focal sizes when the axis of the laser beam is hitting the cross of two filaments.
There is a strong difference, on the other hand, between 50H and 100H cases.
In the former we can identify two channels of low-density plasma, corresponding to the directions along which the laser was scattered (see Figure \ref{fig:scattering} at 0.1 ns for the 50H case).
This quick evaporation of the material leads to a early time for breakout, which we can identify to be 6.4 ns in this last case.
The 100H case is similar to the 50H one, but with a slower propagation of the erosion wave which breaks out of the target at 8.2 ns.

As discussed in Section \ref{sec:exp-results}, the breakout time was measured also experimentally.
We have no information on the precise pointing of the laser for each shot from the experiments and we cannot discriminate among the cases in Figure \ref{fig:focusing-cases}. 
We can anyway reasonably assume that the most probable scenario is the one with the center of the focal spot located at intermediate position between the C and the H cases.
We can therefore identify a maximum and a minimum expected average speeds for the erosion wave, the former corresponding to the H cases, the latter corresponding to the C cases, in accordance with the above discussion.
The obtained velocities are reported in Table~\ref{tab:breakout-speeds}.
\begin{table}[tbp]
    \centering
    \begin{tabular}{ c c c } 
        \hline
        \textbf{Focusing case} & \textbf{Breakout time} & \textbf{Speed} \\ \hline \hline
        50C  & 9.3 ns  & 21.5 µm/ns \\ \hline
        100C & 9.4 ns & 21.3 µm/ns \\ \hline
        50H  & 6.4 ns & 31.3 µm/ns \\ \hline
        100H & 8.2 ns & 24.4 µm/ns \\ \hline
    \end{tabular}
    \caption{The breakout times and the average speeds of the erosion wave obtained in the different cases.}
    \label{tab:breakout-speeds}
\end{table}

\begin{figure*}[tbp]
    \centering
    \begin{subfigure}[b]{0.495\textwidth}
        \includegraphics[width=\textwidth]{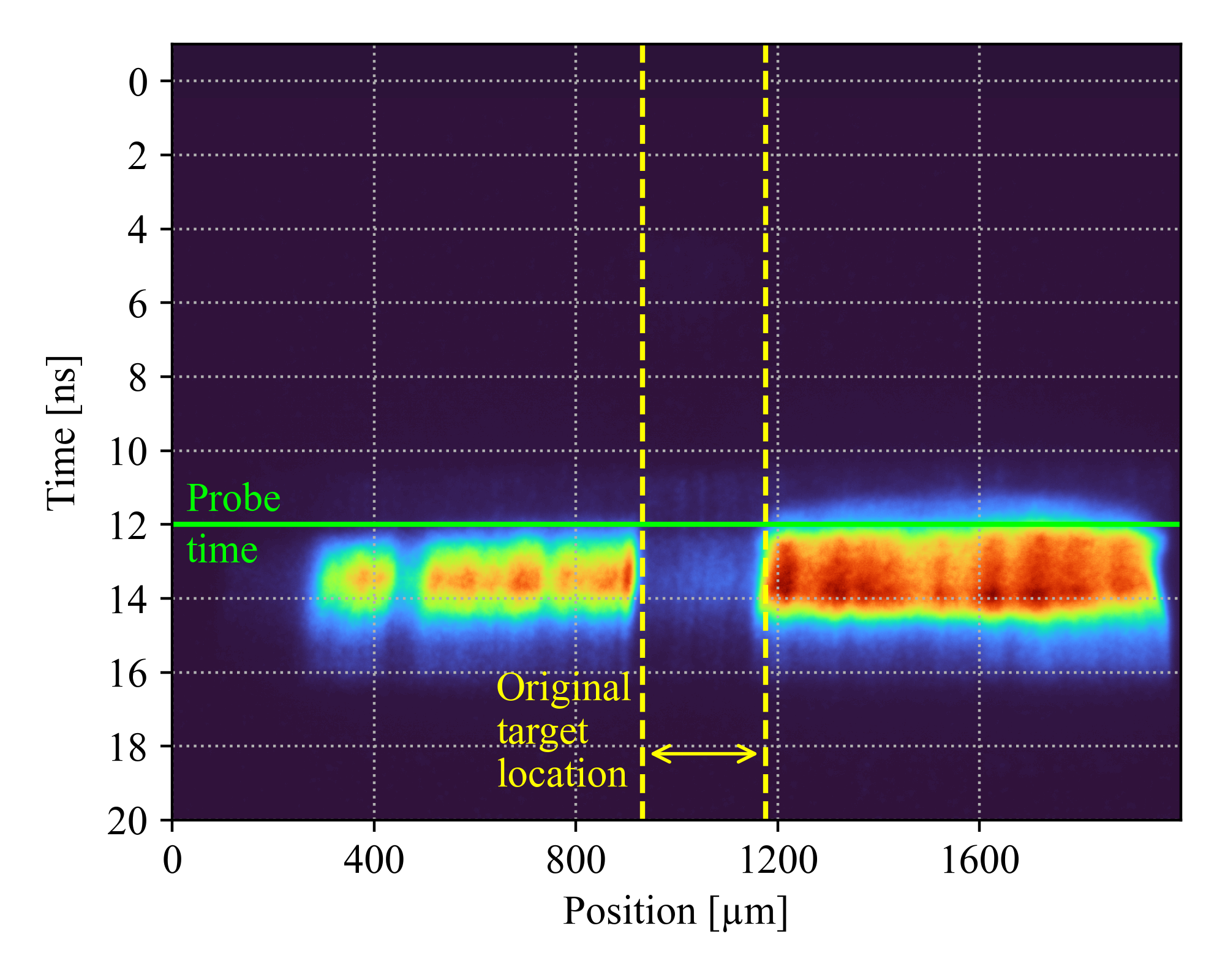}
        \caption{}\label{fig:sample-streak-reference}
    \end{subfigure}
    \begin{subfigure}[b]{0.495\textwidth}
        \includegraphics[width=\textwidth]{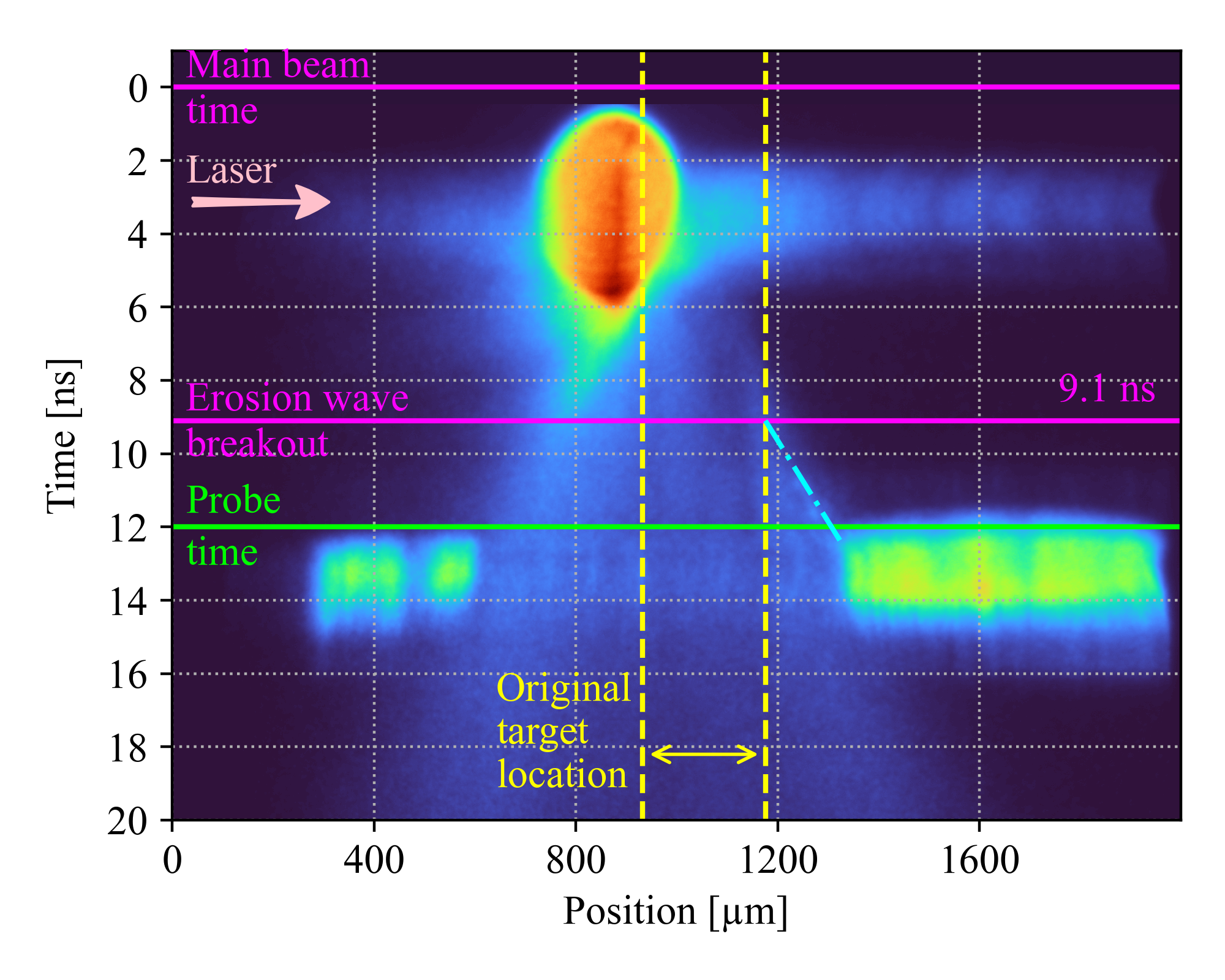}
        \caption{}\label{fig:sample-streak-shot}
    \end{subfigure}
    \caption{A typical streaked image, obtained for the shot 6416. (a) The reference shot made with the only the probe beam, by damping the main one; (b) the streaked image of the shot, where the plasma self-emission is visible. The vertical yellow dashed lines in both images identify the position of the target; the horizontal green line indicates the time of the probe beam; the purple horizontal lines indicate the time of arrival of the main beam on the target (upper line) and the time of erosion wave breakout (lower line); the dotted-dashed cyan line indicates the self-emission of the plasma expanding from the rear side. The main beam in (b) comes from the left of the image.}
    \label{fig:sample-streak}
\end{figure*}

\section{Experimental results} \label{sec:exp-results}

\subsection{Speed of the erosion wave}\label{sec:wave-speed}

The large filament thickness and their large separation did not allow to reach the homogenization of the plasma during the interaction, as the simulations reported in Section \ref{sec:simulations} show.
By using the streaked image from the side of the target, we were able to determine the average speed of the erosion wave into the sample, by following the same strategy used in \citep{Cipriani2018b}.
In Figure \ref{fig:sample-streak} we show a typical example of a streaked image as obtained during the experiments.
To precisely locate the front and rear surface of the target, we performed a reference shot by damping the main beam before entering the interaction chamber, getting Figure \ref{fig:sample-streak-reference}.
In this way we saw the shadow of the target on the streak camera produced on the probe beam and thus locate the actual position of the target in the image.
Figure \ref{fig:sample-streak-shot} shows the streaked image obtained during the irradiation by the main beam.
As discussed in Section \ref{sec:the-experiment}, the probe beam reached the streak to be used as a fiducial.
It is visible in both Figures \ref{fig:sample-streak-reference} and \ref{fig:sample-streak-shot} with a 12 ns delay from the main beam, as indicated by the horizontal green line.
From the streaked image we can identify the instant at which the erosion wave reaches the rear side of the sample, by finding the intersection between the line identifying the rear side of the target and the line along the edge of the plasma self-emission from the rear side of the target, indicated by the cyan line in Figure \ref{fig:sample-streak-shot}.
For the considered shot, this time corresponds to 9 ns after the beginning of the irradiation, resulting in a measured average speed of $27 \pm 8$~µm/ns.
The large error in this value comes from the blur at the self-emission front, a direct consequence of the strong irregularities of the erosion wave-front, also seen in the simulations in Section \ref{sec:simulations}.

\begin{figure}
    \centering
    \includegraphics[width=0.45\textwidth]{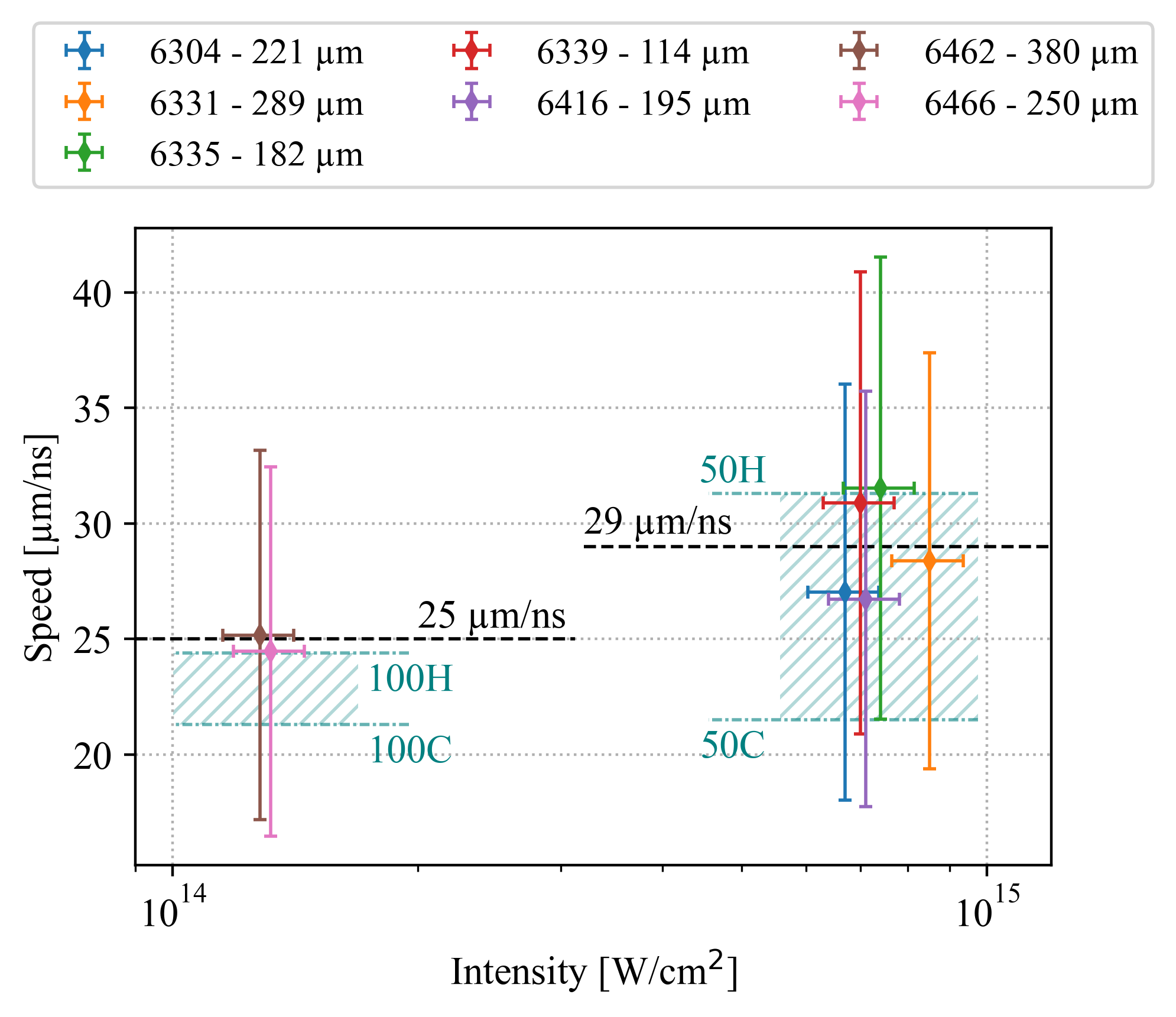}
    \caption{The values of the shock wave speeds calculated from the streaked images. The legend reports the shot number and the thickness of the target for each shot. The shaded areas correspond to the ranges of speeds obtained from the simulations for the two cases of higher and lower intensity, as indicated by the labels (see Table \ref{tab:breakout-speeds}).}
    \label{fig:shock-wave-speeds}
\end{figure}

Figure \ref{fig:shock-wave-speeds} shows the results of this kind of analysis for several shots.
The points in the plot, especially for the higher intensity shots, show a very good stability from shot-to-shot for the speed of the erosion wave.
The measured velocity at about $1.5\times10^{14}$~W/cm$^2$ was of 25~µm/ns, while the one for the shots around $7\times10^{14}$ W/cm$^2$ was of 29~µm/ns on average.
This is within the ranges estimated from simulations, also reported in Figure \ref{fig:shock-wave-speeds}.
In both cases, the speed is generally closer to the one obtained when the center of the focal spot lied in the through hole in the micro-structure, rather than on the crossing of the filaments on the first layer.
This means that the dominant effect with a random pointing of the beam as in our experiments is due, in this range of parameters, to the scattering of the laser light into the structure, leading in general to the effects observed in cases 100H and 50H described in Section \ref{sec:simulations}.

\subsection{Reflected and transmitted light}\label{sec:refl-transm-light}

\begin{figure}
    \centering
    \includegraphics[width=\linewidth]{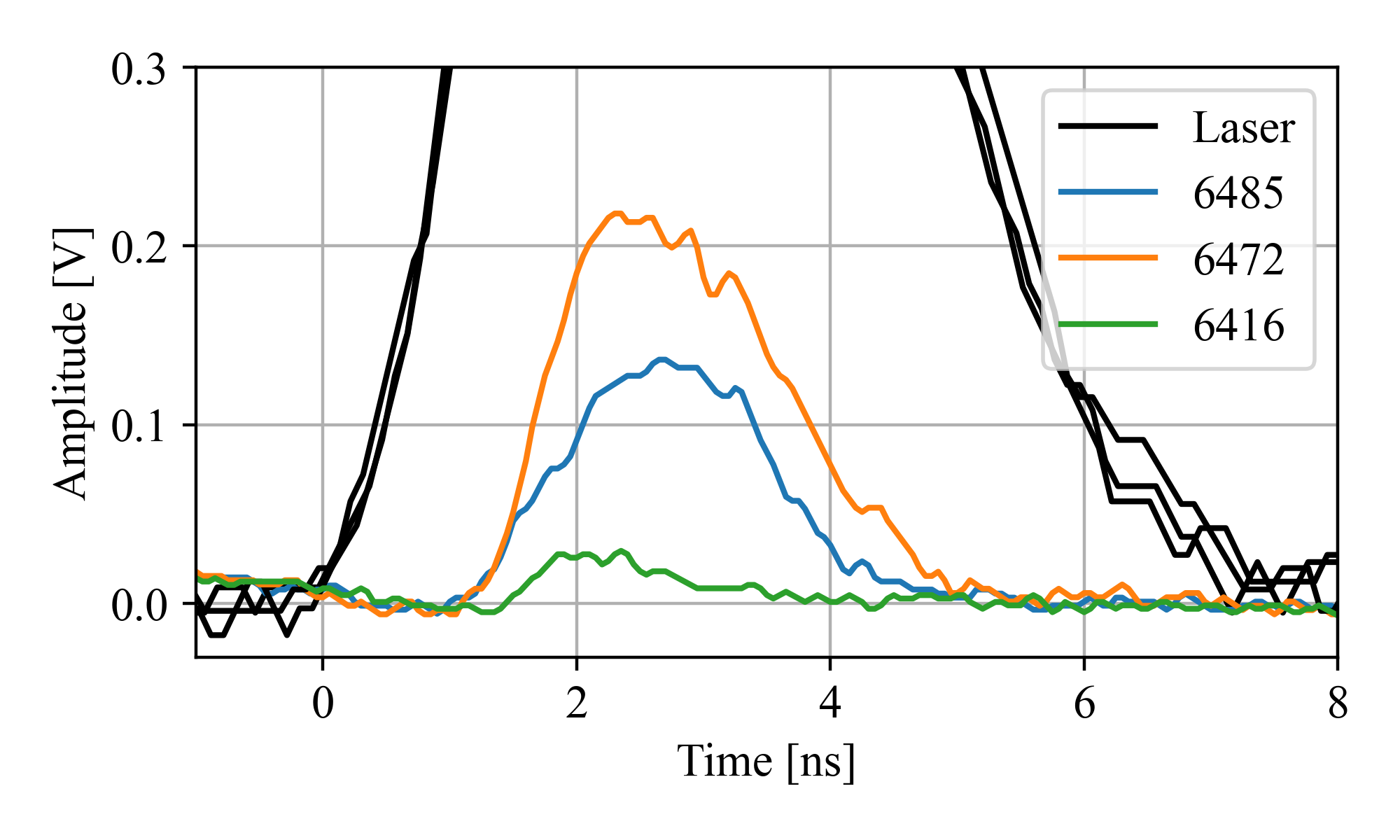}
    \caption{The signals collected by Photodiode R for some selected shots. 
    These are indicated by the colored lines, while the black lines indicate the laser temporal profiles, as measured by Photodiode L.
    The signals are normalized to the maximum of the laser temporal profile.}
    \label{fig:reflected-signals}
\end{figure}

As explained in Section \ref{sec:the-experiment}, we recorded the time-resolved signals from two photodiodes, R and T, collecting the laser light reflected and transmitted by the target, respectively.
For all the shots, we observed a negligible signal from the Photodiode T.
This is consistent with the high average density of the sample and with the simulations.
Figure \ref{fig:reflected-signals} shows three typical signals from Photodiode R recorded during the experiments, whose amplitude has been normalized to the maximum of the signal on Photodiode L, which recorded the light from the main beam, for an appropriate comparison.
The amplitude of the signals had a very wide variability from shot to shot, probably due to a large scattering and diffusion of the laser on the thick filaments and on the strongly inhomogeneous plasma, as also seen from the simulations in Section \ref{sec:simulations}.
Since we collected the reflected light through the focusing lens, some of the light being diffused may have not been directed to the Photodiode R and therefore not detected.
On the other hand, on average, the time duration of reflection was similar, as also seen from the same Figure.

\begin{figure}
    \centering
    \includegraphics[width=\linewidth]{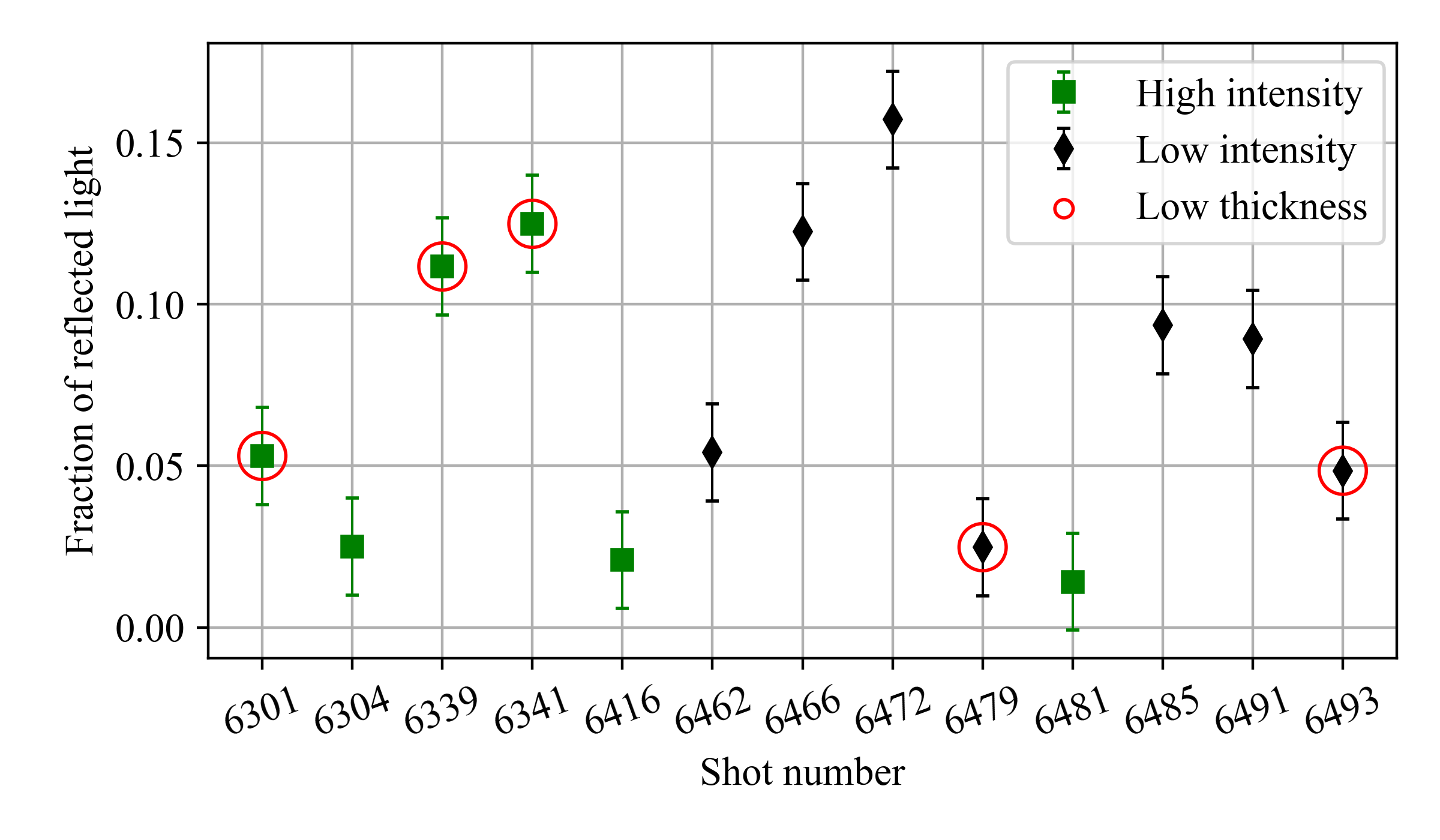}
    \caption{The integrals of the signals from Photodiode R, normalized to the integral of the signal from Photodiode L for better comparison among the different shots. 
    The points labeled with "High intensity" are relative to the shots at about $7\times 10^{14}$ W/cm$^{2}$, while the ones labeled with "Low intensity" had an intensity of $1.5 \times 10^{14}$ W/cm$^2$.
    The red circles indicate the shots in which the target thickness was lower than 150 µm.}
    \label{fig:reflected-integral}
\end{figure}

Figure \ref{fig:reflected-integral} shows the integral of the reflected light signal from Photodiode R normalized to the integral of the signal from Photodiode L.
In the plot we indicate the shots performed at high intensity and the ones at low intensity.
The red circles indicate the shots with a target with a thickness lower than 150 µm, to determine how much the reduced thickness and a potential early erosion of the foam impacted on the reflection of the laser light.
On average, the high-intensity shots tended to be less reflective than the ones at lower intensity.
However, from the integrals we see that there is no evident correlation among the thickness of the target, the laser intensity and the reflectivity.
Since we changed the focal spot size to change the intensity, this is probably related to the large scattering expected when the focal spot size is of the order of the separation between the filaments.

\subsection{Laser-plasma instabilities}\label{sec:LPIs}

As indicated in Section \ref{sec:the-experiment}, a spectrometer was fielded in the experiment, watching the side of the target and collecting the self-emitted light from the plasma.
Some of the shots had an unexpectedly high amplitude in correspondence to the signature of two-plasmon decay instability (TPD), i.e. the self-emission at $2\lambda_L/3$, $\lambda_L$ being the laser wavelength.
In our setup, this corresponds to $\lambda_{2/3} = 703 $ nm.

\begin{figure}[tbp]
    \centering
    \includegraphics[width=\linewidth]{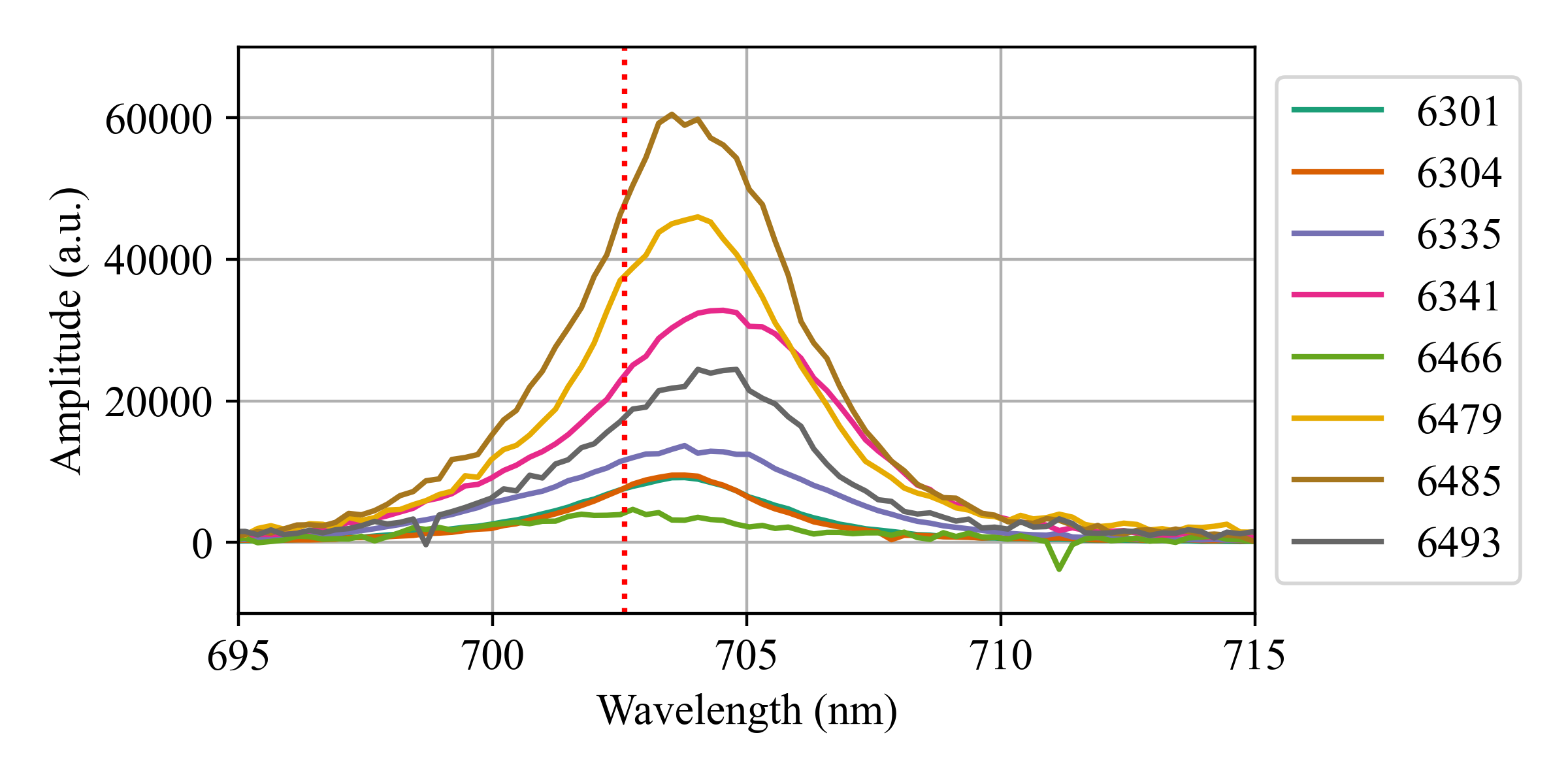}
    \caption{Some of the signals collected by the spectrometer for the $2\lambda_L/3$ self-emission related to the two-plasmon decay instability. 
    In some shots the signal was saturated and are not thus reported.
    The vertical dashed red line corresponds to the exact value of $2\lambda_L/3$ for our experimental conditions.}
    \label{fig:multiple-spectra}
\end{figure}

Figure \ref{fig:multiple-spectra} shows some of the signals collected during the campaign.
The signal amplitude had a large variability in the various shots.
Since the spectrometer was watching the plasma from the side, this may be due to some statistical slight tilting of the target, which may be shadowing the self-emission from the plasma in some cases.
Therefore we cannot use the amplitude as a quantitative estimate of the amount of self-emission and then of the amount of TPD, but qualitatively we know that this instability has been active in our experimental conditions.
Figure \ref{fig:multiple-spectra} shows a clear shift of the emission at $2\lambda_L/3$ and a broadeining.
The shift of the maximum of the signals reported in the Figure ranges from 0.8 to 1.8 nm towards larger wavelenghts, meaning that only the effect from the red plasmon was detectable in our conditions.

\section{Conclusions and future developments}\label{sec:conclusions}

Understanding the physics at the basis of the plasma evolution in foams is very challenging, both from the theoretical and experimental point of view.
In both cases, the randomness of the internal structure of a chemical foams limited for decades our ability to probe the plasma into the material and also to properly simulate it.
One of the main reasons of the actual high attention on 3D-printed foams is the possibility to shape their internal morphology, which is opening new ways of probing the plasma evolution and in the simulation of the fundamental processes.
In fact, simulating the behavior of the target with an ordered structure is much more feasible in an hydrodynamic code than reproducing the evolution of a stochastic net of filaments as in chemical foams \cite{Milovich2021}.

This work represents one of the few numerical and experimental efforts in the literature \cite{Jones2021, Igumenshchev2023, Limpouch2025} devoted to the study of the behavior under irradiation of printed foams.
The structure of the samples used in this work was conceived to enable 3D simulations with a reasonable computational effort, widening the spatial scales at play. 
This allowed to explore the evolution of the plasma in detail.
As discussed in Section \ref{sec:simulations}, in our conditions no shock wave developed, because the homogenization of the foam was never reached.
We rather observed the process of progressive destruction of the material, which we called an erosion wave.
The simulations also show that a regular structure with a large separation between the filaments as with our samples corresponds to a large scattering of the laser in the beginning of the interaction.
This does not only correspond to a fraction of the laser energy being loss through transmission, but also reflects in a peculiar behavior of the plasma, with the formation of zones where the erosion is more efficient, with a less even front of the erosion wave.

We experimentally measured the average speed of the erosion wave, which proved to be in good agreement with the results of the simulations and well reproducible from sample to sample.
This was possible thanks to the shot-to-shot stability of the ABC laser beam and to the high reproducibility of the structure of the material, which was nominally the same over the whole dataset.
On the other hand, other observables, such as the reflected light, were not so reproducible from shot to shot.
This was probably due to the impossibility of an exact pointing of the laser and to the similarity between the spot size (especially at high intensity) with the filament separation.
The amplitude of the $3/2\omega_L$ self-emission related to the two-plasmon decay was also fluctuating, probably because of a non-optimal alignment of the target with the spectrometer.

This work can be extended by using samples with a lower density, of 100 mg/cm$^3$ or below, with thinner filaments.
This will increase the computational cost of the simulations, but it will represent a reliable way of understanding the influence of the morphology on the features of the interaction.
The effects on the plasma dynamics of a reduced spacing between the filaments could be also explored as well.
Experimentally, it will be possible to test the effect of a finer structure on the observables and their stability from shot to shot.
We expect that the erosion wave we discussed will gradually turn to a well defined shockwave as the filament thickness and the inter-filament spacing will be reduced.

\begin{acknowledgements}
This work has been carried out within the framework of the EUROfusion Consortium, funded by the European Union via the Euratom Research and Training Programme (Grant Agreement No 101052200 — EUROfusion).
Views and opinions expressed are however those of the author(s) only and do not necessarily reflect those of the European Union or the European Commission. Neither the European Union nor the European Commission can be held responsible for them.
The software used in this work was developed in part by the DOE NNSA- and DOE Office of Science-supported Flash Center for Computational Science at the University of Chicago and the University of Rochester.
The computing resources and the related technical support used for this work have been provided by CRESCO-ENEAGRID High Performance Computing infrastructure and its staff \cite{9188135}. CRESCO-ENEAGRID High Performance Computing infrastructure is funded by ENEA, the Italian National Agency for New Technologies, Energy and Sustainable Economic Development and by Italian and European research programmes, see http://www.cresco.enea.it/english for information.
This research has been carried out in the framework of the ``Universities' Excellence Initiative'' programme by the Ministry of Education, Science and Sports of the Republic of Lithuania under the agreement with the Research Council of Lithuania (project No. S-A-UEI-23-6). Additional support was received through EU LASERLAB-EUROPE JRA-extension (grant agreement No. 871124, Horizon 2020 research and innovation programme).
\end{acknowledgements}

\section*{Author declarations}
\subsection*{Conflict of interest}
The authors have no conflicts to disclose.

\subsection*{Author contributions}
\textbf{M. Cipriani}: Writing -- original draft (lead); Software (lead); Visualization (equal); Conceptualization (equal); Investigation (equal); Formal analysis (equal); Data curation (equal); Writing -- review and editing (equal).
\textbf{F. Consoli}: Conceptualization (equal); Investigation (equal); Formal analysis (equal); Writing -- review and editing (equal).
\textbf{M. Scisciò}: Conceptualization (equal); Investigation (equal); Formal analysis (equal); Writing -- review and editing (equal).
\textbf{A. Solovjovas}: Writing -- original draft (supporting); Visualization (equal); Resources (equal).
\textbf{I. A. Petsi}: Writing -- review and editing (equal).
\textbf{M. Malinauskas}: Writing -- original draft (supporting); Conceptualization (equal); Visualization (equal); Writing -- review and editing (equal); Resources (equal).
\textbf{P. Andreoli}: Conceptualization (equal); Data curation (equal); Investigation (equal).
\textbf{G. Cristofari}:  Data curation (equal); Investigation (equal).
\textbf{E. Di Ferdinando}:  Data curation (equal); Investigation (equal).
\textbf{G. Di Giorgio}:  Data curation (equal); Investigation (equal).

\section*{Data Availability}

The data that support the findings of this study are available from the corresponding author upon reasonable request.

\appendix

\section{Target manufacturing}\label{app:manufacturing}

To produce the micro-structures, we used two different manufacturing systems: the Asiga Pico 2 UV 3D stereolithography printer operating at 385 nm central wavelength and a LDW MPL3DL utilizing the 515 nm wavelength. 
The full experimental set-up is presented in Fig. 1. 
The holders for the structures were made by using the table-top 3D UV printer and Asiga Plasgray photo-resin, with a printing time of approximately 12 minutes. 
The samples were then dissolved in the isopropyl alcohol to remove all non-polymerized photo-resin.
To fabricate the micro-structures we used the hybrid pre-polymer SZ2080\texttrademark\ with the LDW system. 
In general, there are two options to prepare the solution for LDW manufacturing, i.e. to mix SZ2080\texttrademark\ with 1\% of the photo-initiator Irgacure 369 or to use clean pre-polymer SZ2080\texttrademark. 
The decision not to use the photo-initiator was made reduce the amount of toxic ingredients in the process. It was recently demonstrated the possibility for green 3D printing, avoiding the mixing the photo-initiators\cite{24lam}. In addition, the usage of clean SZ2080\texttrademark\ allows to avoid unwanted particles from the photo-initiator which could result in bubble formation leading to micro-explosions during the formation of the structure.
We also reduced the risk of bubble formation by cleaning the glass printing plate with Isopropyl alcohol for a few minutes.
To print the micro-structure inside the already printed holder, we mounted the latter on a cover glass, and then drop-cast SZ2080\texttrademark\ pre-polymer on it. 
After that, we annealed the sample for 100 minutes at 90 degrees Celsius. 
During this process, the sample holder might start to curl due to an increase of temperature, potentially causing the detachment of the holder from the glass.
In this case, printing it is not still possible because of a severe degradation of the quality of the final product and the sample has to be discarded.
For this reason, 8 holders were annealed at once with SZ2080\texttrademark\ pre-polymer to increase the success rate with a single printing. 
We also kept the wall thickness of the holder in the CAD model to be at least 400 µm, to avoid any deformation during drop-casting.
For the LDW manufacturing, we used an Yb:KGW laser (Pharos, Light Conversion, Ltd.) operating at 180 fs pulse duration, with 200 kHz pulse repetition rate and 1030 nm central wavelength. 
In addition, we used a BBO crystal to generate the second harmonic for the micro-fabrication process, for the wavelength to allow two-photon absorption in the SZ2080\texttrademark~resin.
We focused the laser beam with a 0.8 numerical aperture objective. 
We modified the CAD model for the micro-structure for every sample to compensate any variation in the distance between the parts of the holder that had to support the micro-structure with respect to the original one.
In the reality, this distance was found to vary from about 490 µm to 520 µm and we had to consider the log-pile to extend about 50 µm more in the end, to be sure that the LDW printing was started and ended on the holder supports, for better adhesion.
The printing of the log-piles took from 60 minutes to about 100 minutes, depending on the structure height and the writing speed. 
The most favourable laser power for this process was found to be between 0.2 and 0.3~mW (corresponding to 0.298 and 0.447 TW/cm$^2$ intensities at the focal spot within the sample volume), the power region where SZ2080\texttrademark\ polymerizes without micro-explosions.
When the printing was completed, we cleaned the sample in a Petri glass filled with pentanon. 
We left the sample in the Petri glass for at least half an hour, to allow the sample holder with the micro-structure to detach from the glass.
This time has to be carefully controlled, to prevent the pentanon to dissolve too much the sample holder, so that it won't stand straight.

\begin{center}\textbf{References}\end{center}

\bibliography{EXP_ABC-2PPFoams, CRESCObib, VULRCbib}

\end{document}